# Double Generalized Linear Model Reveals Those with High Intelligence are More Similar in Cortical Thickness

Qi Zhao[1], Lingli Zhang[2], Chun Shen[3], Jie Zhang[3], Jianfeng Feng[3]

1. School of Mathematical Sciences, Fudan University, Shanghai, China,
2. Shanghai Children's Medical Center, Shanghai Jiao Tong University School of Medicine, Shanghai, China,
3. Institute of Science and Technology for Brain Inspired Intelligence, Fudan University, Shanghai, China

## Abstract

Most studies indicate that intelligence (g) is positively correlated with cortical thickness. However, the interindividual variability of cortical thickness has not been taken into account. In this study, we aimed to identify the association between intelligence and cortical thickness in adolescents from both the group's mean and dispersion point of view, utilizing the structural brain imaging from the Adolescent Brain and Cognitive Development (ABCD) Consortium, the largest cohort in early adolescents around 10 years old. The mean and dispersion parameters of cortical thickness and their association with intelligence were estimated using double generalized linear models (DGLM). We found that for the mean model part, the thickness of the frontal lobe like superior frontal gyrus was negatively related to intelligence, while the surface area was most positively associated with intelligence in the frontal lobe. In the dispersion part, intelligence was negatively correlated with the dispersion of cortical thickness in widespread areas, but not with the surface area.

These results suggested that people with higher IQ are more similar in cortical thickness, which may be related to less differentiation or heterogeneity in cortical columns.

## Introduction

Intelligence is the most heritable of all cognitive measures, and develops rapidly during adolescence. General intelligence ($g$) can be defined as the weighted sum of fluid ($gF$) and crystallized ($gC$) intelligence. $gF$ refers to reasoning ability, and is related to working memory(Deary and Caryl 1997); while $gC$ is sometimes described as verbal ability, and is more dependent on accumulated experience.

Gray matter is closely related to intellectual ability, and it has been shown that intelligence and cortical thickness are partly associated through shared genes ([Brans, Kahn et al. 2010](#)). Recent neuroimaging studies have investigated the correlation between human intelligence and cortical thickness. For example, in young adults (20.9 $\pm$ 2.9 years), *g* was strongly related to cortical thickness in regions of the temporal lobe ([Choi, Shamosh et al. 2008](#)). However, a longitudinal study ([Shaw, Greenstein et al. 2006](#)) described that there is a developmental shift from a predominantly negative correlation between intelligence and cortical gray matter thickness in early childhood (3.8 to 8.4 years) to a pronounced positive correlation in late childhood (8.6 to 11.7 years) in superior frontal gyri, medial prefrontal cortex, middle and orbitofrontal cortices. As the neuroanatomical expression of intelligence in children and adolescents is dynamic, it might help better illustrated this relationship in the participants with narrow age distribution to avoid developmental periods in which the brain changes radically.

Previous studies focusing on the correlation between brain morphology and IQ only explored the association between group mean of neuroimaging measures and IQ, and did not model the interindividual variability, or dispersion in the group, which carry relevant information regarding gene-environment interactions related to the individual sensitivity to environmental and genetic perturbation([Alnaes, Kaufmann et al. 2019](#)).Therefore, we first using double generalized linear model (DGLM) to model the variability between brain cortical thickness and intelligence in preadolescents. We utilized the structural brain magnetic resonance imaging from a large sample of the ABCD Study (Casey, Cannonier et al. 2018), with an narrow age span (around 10 years old) and identify the association both from correlation and variability using DGLM, which allows the mean and dispersion to be modelled simultaneously in a generalized linear model context(Smyth and Verbyla 1999). Meanwhile, the association between intelligence and surface area were also analyzed to compare with thickness results.

**Results**

**Study samples**

Our study included 10,666 subjects after quality control for FreeSurfer v5.3.0 (N=11,076) and remove subjects with lack cognitive score. The samples include 22 sites; sex, age and mean thickness information were listed in the **Table1**. We also analyzed the correlation between different scores and covariables (**FigS1** in Supplementary).

**Mean model results**

In the mean model, we model the relationship between cortical thickness and $g$ in the mean part. Cortical thickness were predominantly negatively correlated with $g$ in some frontal lobe and limbic lobe (**Fig1-2**), including

G_front_sup, G_front_middle, S_suborbital, right S_orbital_med-olfact, G_and_S_cingul-Ant,S_pericallosal and S_temporal_transverse. Meanwhile, $g$ were predominantly positively correlated with some frontal lobe (S_precentral-inf_part, G_and_S_subcentral, G_orbital), temporal regions (S_oc-temp_med_and_Lingual, S_temporal_inf, S_temporal_sup, Pole_temporal), parietal lobe (G_and _S_postcentral), limbic lobe (G_oc-temp_med-parahip) and major divisions (G_Ins_lg_and_S_cent_ins, S_calcarine, S_parieto_occipital). However, cortical surface area were globally positively associated with $g$, $gC$ and $gF$ (**FigS4** in Supplementray).

The prominent brain regions correlates were almost symmetrically distributed in $g$, $gF$ and $gC$. However, $gC$ were predominantly more associated with cortical thickness than $gF$ in the mean model ($gC$: 109 regions > $gF$: 65 regions). Moreover, the part that $gC$ differ from $gF$ was mostly in left hemisphere (left: 17, right: 7). For example, left S_circular_insula_sup, left Pole_occipital, left S_temporal_inf, left G_front_inf-Orbital, left S_front_inf and left Lat_Fis-ant-Horizont.

**Dispersion model results**

In the dispersion model, we model the relationship between cortical thickness and $g$ in the dispersion/variance part. Interestingly, $g$, $gC$ and $gF$ were all negatively associated with cortical thickness dispersion (**Fig3-4**). However, $g$, $gC$ and $gF$ were not associated with surface area dispersion.

$gF$ correlated more regions than $gC$ in the dispersion model ($gF$: 25 regions > $gC$: 11 regions). $gF$ was negatively associated with dispersion in frontal lobe (G_and_S_frontomargin, S_front_middle, S_precentral-inf-part and S_suborbital), parietal lobe (S_parieto_occipital, G_and_S_postcentral, G_and_S_intrapariet_and_P_trans, G_and_S_postcentral, G_precuneus, S_subparietal), occipital lobe (S_oc_middle_and_Lunatus, S_oc_sup_and_transversal, G_oc-temp_lat-fusifor, S_oc-temp_lat, S_oc_sup_and_transversal, S_oc_middle_and_Lunatus, G_occipital_sup, S_oc-temp_med_and_Lingual), S_pericallosal, G_and_S_cingul-Ant and circular sulcus of the Insula.    $gC$ was negatively associated with dispersion in temporal lobe (Pole_temporal, S_collat_transv_ant, S_temporal_inf), limbic lobe (G_and_S_cingul-Ant, G_and_S_cingul-Mid-Ant), G_precuneus, G_orbital, S_oc_middle_and_Lunatus and S_parieto_occipital. Besides, $gC$ was more associated with dispersion in left-hemi cortex regions (left: 7, right: 4), and $gF$ are more associated with dispersion in right-hemi cortex regions (left: 12, right: 13).

**Cross-Validated Elastic Net Regression results**

We set the α value to 0.5 to take advantage of the relative strengths of the two above regression approaches, providing a no sparse solution with low variance among several correlated independent variables. The cortical thickness, surface area respectively accounted for about 10%, both accounted for about 14% of the total variance of cognition total composite score age-corrected standard score. This $R^2$ was significantly higher than expected due to chance (P <0.001, compared with $R^2$ from 500 randomly generated elastic net regressions). Correlations between actual standard score versus predicted cognitive standard scores, averaging across 10 folds of the cross-validation, were $gF$ r = 0.25, $gC$ r = 0.29 and $g$ r = 0.32 using cortical thickness(Fig.5), $gF$ r = 0.24, $gC$ r = 0.31 and $g$ r = 0.31 using cortical surface area and $gF$ r = 0.30, $gC$ r = 0.36 and $g$ r = 0.37 using both cortical thickness and surface area (FigS5 in the supplementray).

**Discussion**

The most important finding of the present work was that, higher intelligence was associated with lower interindividual heterogeneity in cortical thickness, for example in parieto-occipital sulcus, anterior part of the cingulate gyrus and sulcus. This means higher IQ population has lower variation than lower IQ population in some regions (FigS3), possibly reflecting higher IQ population are more similar in brain structure than lower IQ population. However, intelligence was not associated with brain heterogeneity in cortical surface area. Cortical thickness and surface area were both highly heritable but were essentially unrelated genetically (Panizzon, Fennema-Notestine et al. 2009).    From neuronal point of view, cortical thickness is associated with radial neuronal migration and number of neurons, dendritic arborizations, and glial support in cortical columns, while surface area is related to tangential neuronal migration and captures of mini-columnar units in the cortex (Chenn and Walsh 2003, Rakic 2009, Rakic, Ayoub et al. 2009, Tadayon, Pascual-Leone et al. 2019). Therefore, our results suggested that in the early stages of development (around 10 years old), those with higher IQ had less differentiation in cortical columns, and that their brain morphology developed following a similar trajectory that leads to higher IQ. Interestingly, these areas overlap substantially with the default network. A recent review paper proposed that distributed association networks in default network are supported by anatomical connectivity(Buckner and DiNicola 2019). Thus, this finding might reveal the underlying relation between morphology and default network in the development process.

From the dispersion model, we further found significant asymmetry of left and right hemispheres in terms of the correlation of their cortex thickness with gF/gC. gC is associated with dispersion in the left hemisphere regions; the fluid intelligence is associated with dispersion in right hemisphere

Regions. This phenomenon may reflect that gF/gC can be studied from another perspective using the DGLM model.

For the mean model, we found that higher intelligence was associated with a decrease in thickness in frontal lobe, but an increase in other areas like calcarine sulcus, lingual sulcus, parahippocampal gyrus and central sulcus, which means higher performance was associated with cortical thickness related to working memory, attention, and visio-spatial processing. Interestingly higher intelligence was associated with increase in surface area in almost the whole brain, most prominently in frontal cortex.

The negative correlation between thickness and IQ in frontal lobe was largely in line with recent studies showing cortical gray matter thinning in anterior and superior frontal areas was associated with superior arithmetic performance to 9- and 10-year-old children (Chaddock-Heyman, Erickson et al. 2015). Shaw's longitudinal study showed that superior intelligence clusters demonstrated a marked increase in cortical thickness peaking in superior frontal gyri at around 11 years old, later than average intelligence group (Shaw, Greenstein et al. 2006). This provided an explanation for the negative correlation between thickness and IQ in frontal lobe. Interestingly, the most significant positive correlation between surface area and IQ was also in frontal lobe. Considering the theories that the first step in the evolutionary ascent of the human cerebral cortex is its enlargement, which occurs mainly by expansion of the surface area without a comparable increase in its thickness(Rakic 2009). Taken together, these results suggested that the frontal lobe surface area enlarge at first, and then thickness increases later for preadolescents with higher IQ.

The morphological correlates of subitems of IQ revealed significant difference between $gC$ and $gF$. For $gC$, its two cognitive domain scores, Picture Vocabulary and Oral Reading Recognition task scores exhibited very similar patterns of association with cortical thickness (Fig S6). However, the subitems in gF had different associated with cortical thickness. Therein, working memory are mostly associated with cortical thickness, picture sequence memory, cognitive flexibility (Dimensional Change Card Sort Task) and flanker are less associated with cortical thickness. Pattern comparison processing speed test is negative associated with cortical thickness in the right hemisphere anterior cingulate gyrus and sulcus, parieto-occipital sulcus and temporal inferior sulcus, which all distributed in default network related regions (FigS7).

Using cortical thickness and surface area accounted for about 14% of the total variance of cognition total composite score age-corrected standard score, more than using both respectively. It means cortical thickness and surface area contribute different aspects to g. The findings, based on harmonized analysis protocols for all included data sets, were robust to strict procedures for removing outliers and quality assessment and multisite case-control differences cannot be explained by scanning site. Excluding total brain thickness/area as a cofactor in the model did not influence the association between gF/gC and cortical thickness.

**Methods**

**Samples**

The participants were recruited by the ABCD Study Release 2.0.1 after quality control for neuroimaging data and behavioral tests with an age span between 108-131 months (around 9-10 years old). The ABCD Consortium used NIH Toolbox Cognition battery (NIHTB-CB) composite scores(Luciana, Bjork et al. 2018), which include a Total Score Composite, a Crystalized Intelligence Composite (The Toolbox Picture Vocabulary Task and The Toolbox Oral Reading Recognition Task) and a Fluid Intelligence Composite (The Toolbox Pattern Comparison Processing Speed Test, The Toolbox List Sorting Working Memory Test, The Toolbox Picture Sequence Memory Test, The Toolbox Flanker Task, The Toolbox Dimensional Change Card Sort Task)(Akshoomoff, Beaumont et al. 2013). These composite scores also show good test–retest reliabilities in both children and adults as well as validity in children(Akshoomoff, Beaumont et al. 2013, Heaton, Akshoomoff et al. 2014) and highly related (r=0.89) with scores measured with WAIS-IV(Heaton, Akshoomoff et al. 2014).

Cortical thickness are measured using FreeSurfer 5.3.0 under Destrieux atlas, which include 148 regions. Multiple linear regression models were employed to model the relationship between brain cortical thickness and three cognition scores, separately. Although the age span is narrow, intelligence is significantly correlated with age. Therefore, age-corrected standard scores were used and meanwhile, age, gender and site were considered as nuisance variables in the models.

**Statistical Analysis**

Statistical analyses of demographic data and test scores were conducted using R software. The mean and variability parameters of cortical thickness and their association with intelligence were estimated using double generalized linear models(DGLM) (Efron 1986, Smyth 1989). Before this, age, sex and site were regressed as nuisance variables using generalized additive model(GAM) (Diederich 2007). Then, DGLM iteratively fit a generalized linear model of the mean parameter and a second generalized linear model of the variability parameter on the deviance of the first model and. Cortical thickness statistic map(t statistics) are submitted to correct for multiple comparisons using false discovery rate (FDR)

correction(<u>Benjamini and Hochberg 1995</u> ) and the brain regions with corrected p value less than 0.05 would survive. Finally, elastic net regression was employed to cortical thickness for predicting three kinds of intelligence.

**Generalized additive Model** (GAM): In order to correct the data for site, age, sex effects, we ran generalized additive models on each ROI analyses using the following model:

$$Y \sim s(Age) + Sex + Scanner.$$

Where $Y$ represents cortical thickness in each brain regions, s(.) is a smooth function, estimated from the data.

**Double Generalized Linear Model** (DGLM): DGLM fitted using the following model for both the mean and dispersion part. Modeling the dispersion is important for obtaining correct mean parameter estimates if dispersion varies as a function of the predictor, and allows for systematic investigation into factors associated dispersion in observations.

Mean model: $m_i = \mu + Age\beta_{age} + Sex\beta_{sex} + IQ\beta_{cs} \; i = 1, 2, \dots, N$

Dispersion model: $v_i = v + Age\gamma_{age} + Sex\gamma_{sex} + IQ\gamma_{cs} \; i = 1, 2, \dots, N$

Here, we assume $Y$ is cortical thickness regressed nuisance variables. It follows a normal distribution with expectation $m_i$ and variance $\sigma_i^2$, and $\sigma_i$ is also a function rather than a constant like $m_i$. All $\beta$ are the parameters to be estimated. N is the number of brain regions. For a more intuitive explanation of the model, Figure S2 shows a general view of relationship between different kinds of data distribution and DGLM.

**Cross-Validated Elastic Net Regression**(Zou and Hastie 2005)*: We used elastic net to test whether cortical thickness can predict different kind of intelligence across subjects. Elastic net enables data-driven regression analysis by enforcing sparsity of regression output values (i.e., reducing the number of final $\beta$ regression values). In other words (Casey, Cannonier et al. 2018), it provides automatic variable selection by removing all independent variables not predicted dependent variable. We normalized all input data:

$$\bar{X} = \frac{X - mean(X)}{\max(X) - \min(X)}$$

This resulted in variables, x, with values between 0 and 1. The elastic net equation is then written as

$$\hat{\beta}_0, \hat{\beta} = \arg\min_{\beta_0, \beta} \sum_{i=1}^{n} \left( y_i - \beta_0 - \sum_{j=1}^{p} \beta_j X_{ij} \right)^2$$

$$+ \lambda \sum_{j=1}^{p} \frac{1}{2}(1-\alpha)\beta_j^2 + \alpha|\beta_j|$$

This is a doubly penalized regression model using both LASSO and Ridge regression. $\alpha$ sets the degree of mixing between ridge regression and lasso. Meanwhile, $\beta$ is the shrinkage parameter. When $\beta = 0$, no shrinkage is performed.

**Conclusion**

Ongoing efforts are attempting to account for brain cognitive function and brain morphology. Herein we report that intelligence appears to be associated with widespread increased mean differences and decreased heterogeneity in cortical thickness. The results seem to support the notion that cognitive function has high heterogeneity. Subjects with high IQ have lower heterogeneity in cortical thickness in widespread brain areas. Together these findings warrant future longitudinal studies that cortical thickness contributing to neurobiological heterogeneity.

**Table 1.** Demographic and background characteristics of ABCD samples among 22 sites

| Site | Count | Sex(female/all) | Age [SD] | Mean Thickness [SD] |
|------|-------|-----------------|----------|---------------------|
| 1 | 345 | 0.48 | 118.58[7.62] | 2.79[0.10] |
| 2 | 529 | 0.46 | 120.91[7.51] | 2.81[0.08] |
| 3 | 602 | 0.47 | 118.34[7.42] | 2.8[0.08] |
| 4 | 643 | 0.49 | 117.66[7.82] | 2.69[0.10] |
| 5 | 360 | 0.50 | 118.72[7.41] | 2.8[0.09] |
| 6 | 559 | 0.50 | 119.23[7.16] | 2.82[0.09] |
| 7 | 325 | 0.46 | 118.38[7.52] | 2.79[0.09] |
| 8 | 265 | 0.46 | 119.72[7.42] | 2.69[0.09] |
| 9 | 392 | 0.49 | 119.38[7.36] | 2.78[0.09] |
| 10 | 621 | 0.48 | 118.19[7.57] | 2.69[0.09] |
| 11 | 437 | 0.49 | 117.68[7.63] | 2.79[0.09] |
| 12 | 564 | 0.49 | 118.28[7.4] | 2.78[0.09] |
| 13 | 574 | 0.50 | 117.45[7.29] | 2.69[0.10] |
| 14 | 526 | 0.46 | 122.01[6.79] | 2.82[0.09] |
| 15 | 366 | 0.47 | 118.55[7.34] | 2.76[0.10] |
| 16 | 990 | 0.45 | 118.54[7.88] | 2.83[0.08] |
| 17 | 530 | 0.50 | 117.64[7.57] | 2.81[0.10] |
| 18 | 306 | 0.46 | 119.44[7.57] | 2.68[0.10] |
| 19 | 486 | 0.52 | 120.66[6.62] | 2.78[0.11] |
| 20 | 662 | 0.50 | 120.69[5.86] | 2.81[0.09] |
| 21 | 551 | 0.44 | 118.78[7.53] | 2.78[0.09] |
| 22 | 33 | 0.58 | 122.55[6.49] | 2.67[0.10] |

**A) Total Intelligence mean model**

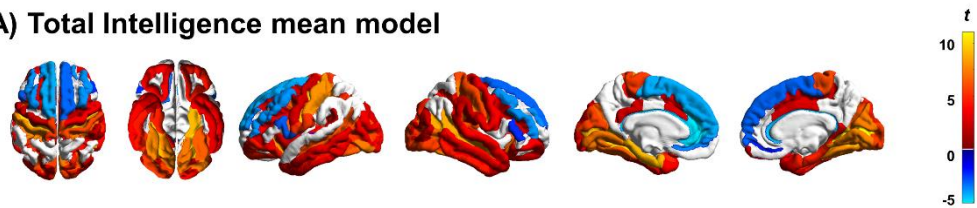

**B) Crystallized Intelligence mean model**

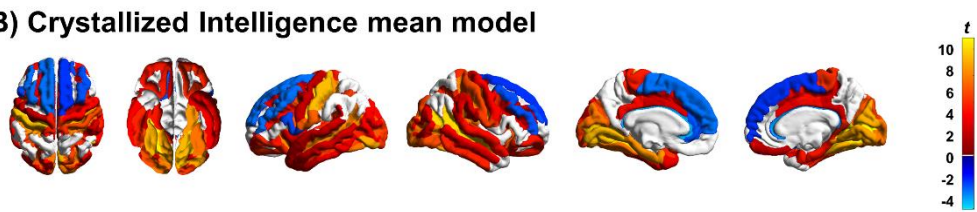

**C) Fluid Intelligence mean model**

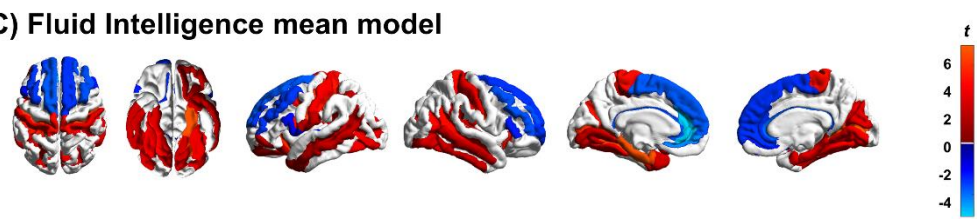

**Fig 1. The mean model results of regions of interest, which the cortical thickness significantly**

**correlation with *g*, *gC* and *gF*.** Cortical regions of interest which p value through FDR (0.05) correction were shown based on Destrieux atlas.

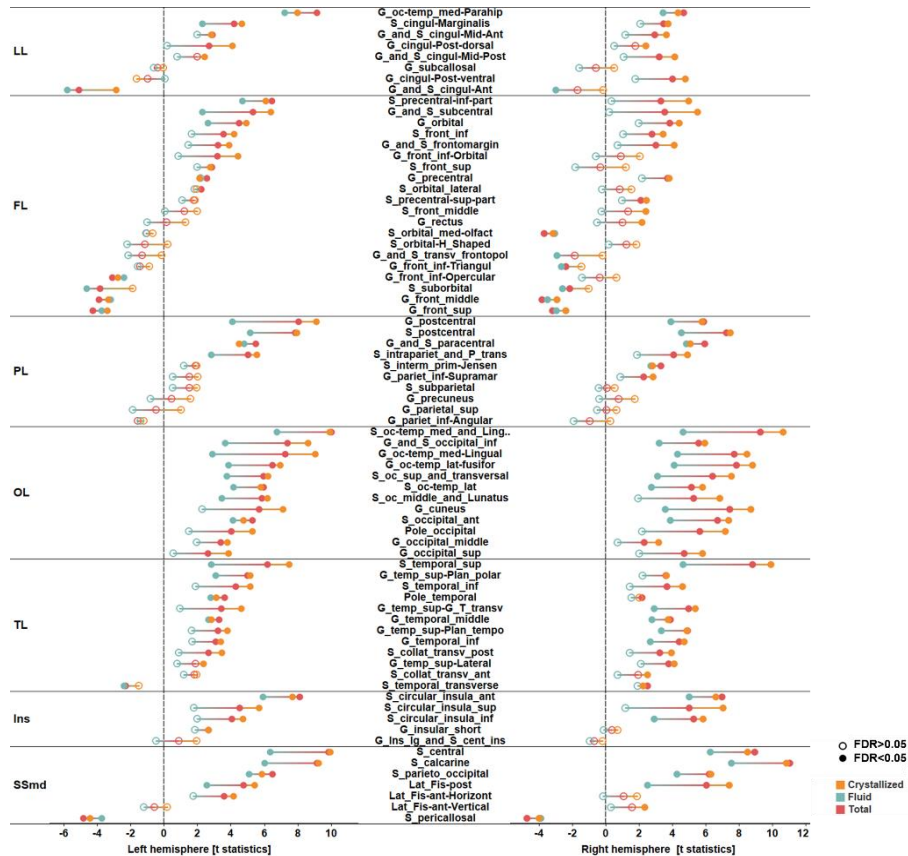

**Fig 2.** The general correlation between the three intelligence indicators and region thickness of cortical regions based on Destrieux atlas. The atlas are further broken down into limbic lobe and sulcus (LL), frontal lobe and sulcus(FL), temporal lobe and sulcus(TL), parietal lobe and sulcus (PL), occipital lobe and sulcus(OL), insular cortex(Ins) and sulci /spaces major divisions(SSmd).

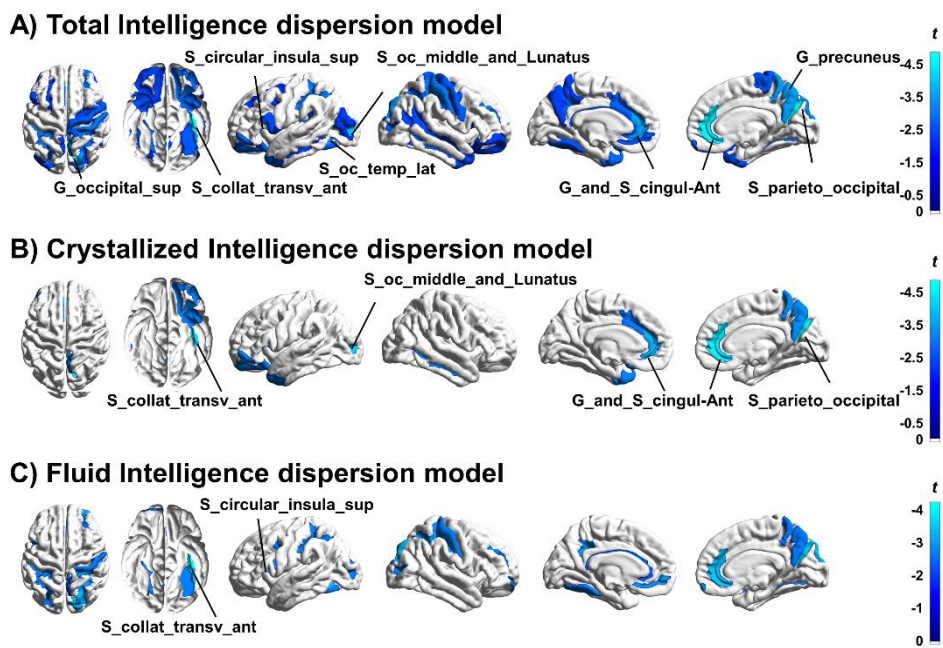

**Fig 3. The dispersion model results of regions of interest, which the cortical thickness significantly correlation with total, crystallized and fluid intelligence.** Cortical regions of interest which p value through FDR(0.05) correction were shown and p values through Bonferroni(0.05) correction were tagged based on Destrieux atlas.

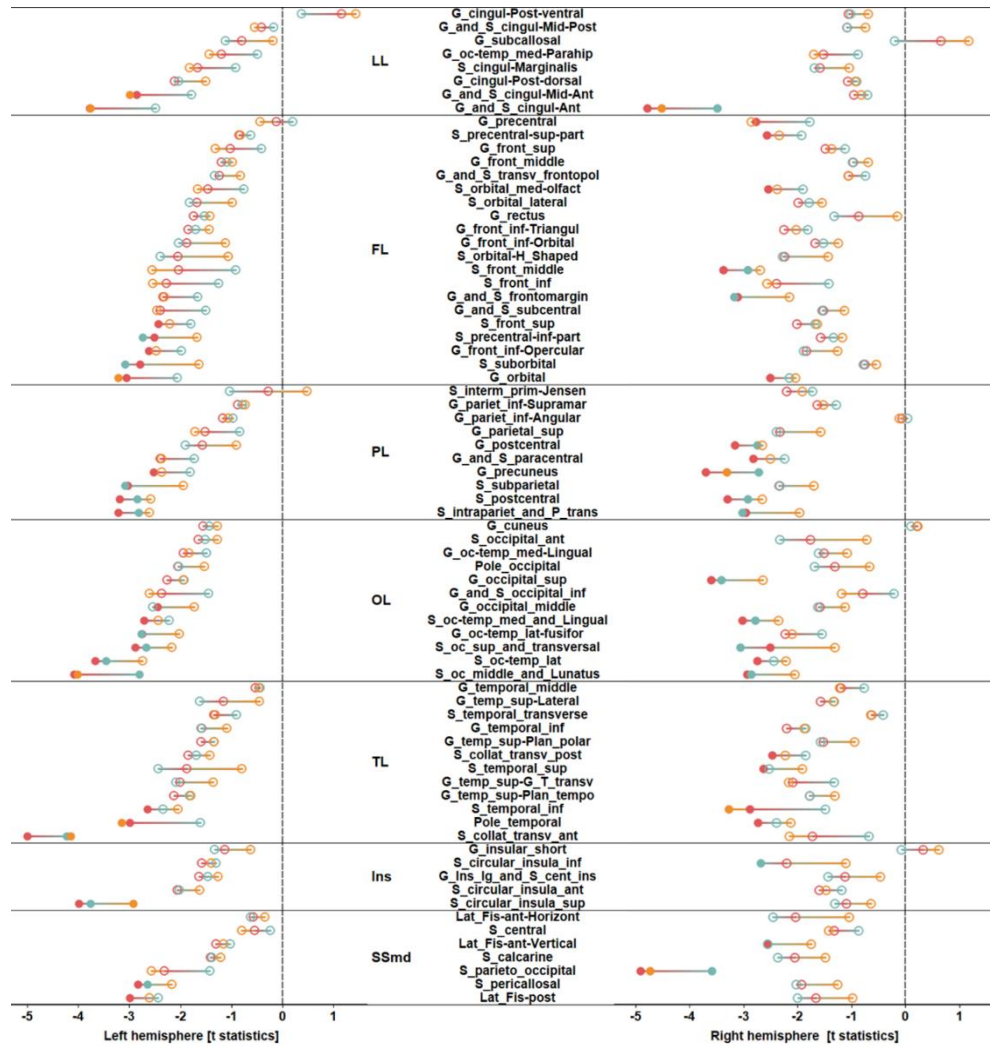

**Fig 4.** The variability correlation between the three intelligence scores and cortical thickness of ROI based on Destrieux atlas.

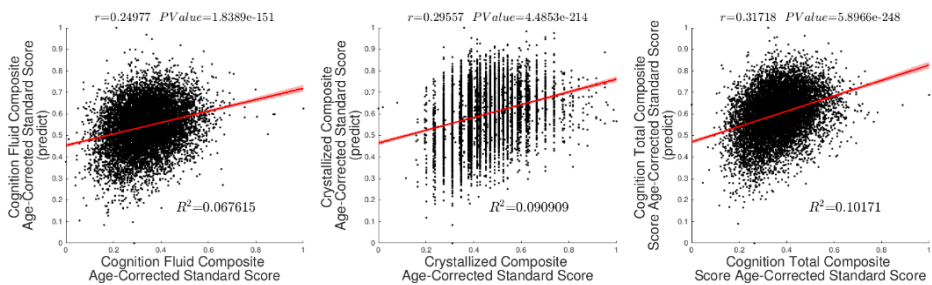

**Fig 5.** Scatterplots between elastic net predicted and normalized intelligence scores (y axes) and original normalized values (x axes). Each dot is a single sample, and dashed lines denote the best linear fit between predicted and normalized intelligence scores.

## Supplementary

ABCD Youth NIHTB-CB Summary Scores include two types of scores: age corrected scores and fully corrected T-Score. Age-corrected scores compare the score of the test-taker to others of the same age. For children, normative scores are provided separately for each year of age to consider expected developmental changes. These are presented as Standard Scores (mean=100, SD=15). Fully Corrected T-Scores (mean = 50, SD = 10) compare the score of the test-taker to those in the NIH Toolbox nationally representative normative sample, while adjusting for key demographic variables. These variables include age, gender, race/ethnicity and educational attainment (for ages 3-17, parent's education is used). All seven of the NIHTB-CB tests were included in this study. This resulted in two measures of crystallized abilities (the NIHTB Picture Vocabulary Test and Oral Reading Test), as well as five measures of fluid abilities: the NIHTB Dimensional Change Card Sort (DCCS) Test of Executive Function-Cognitive Flexibility, NIHTB Flanker Test of Executive Function- Inhibitory Control and Attention, NIHTB Picture Sequence Memory Test of Episodic Memory, NIHTB List Sorting Working Memory Test, and NIHTB Pattern Comparison Processing Speed Test.

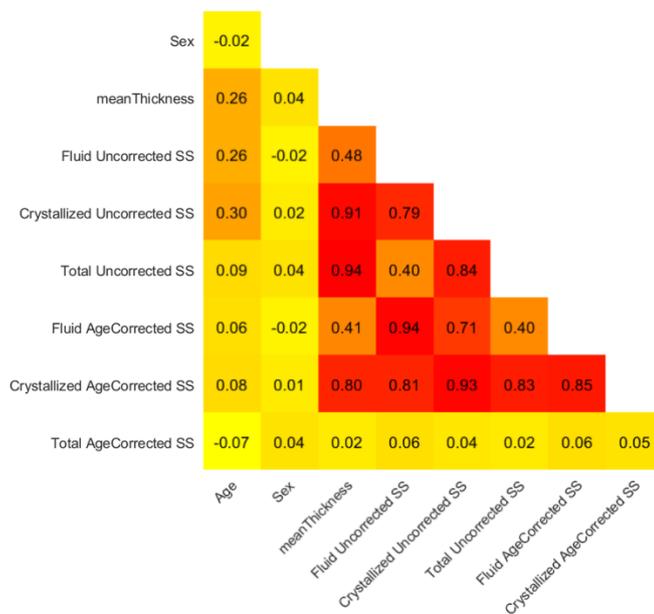

**Figure S1. The correlation between different TB Summary Scores**

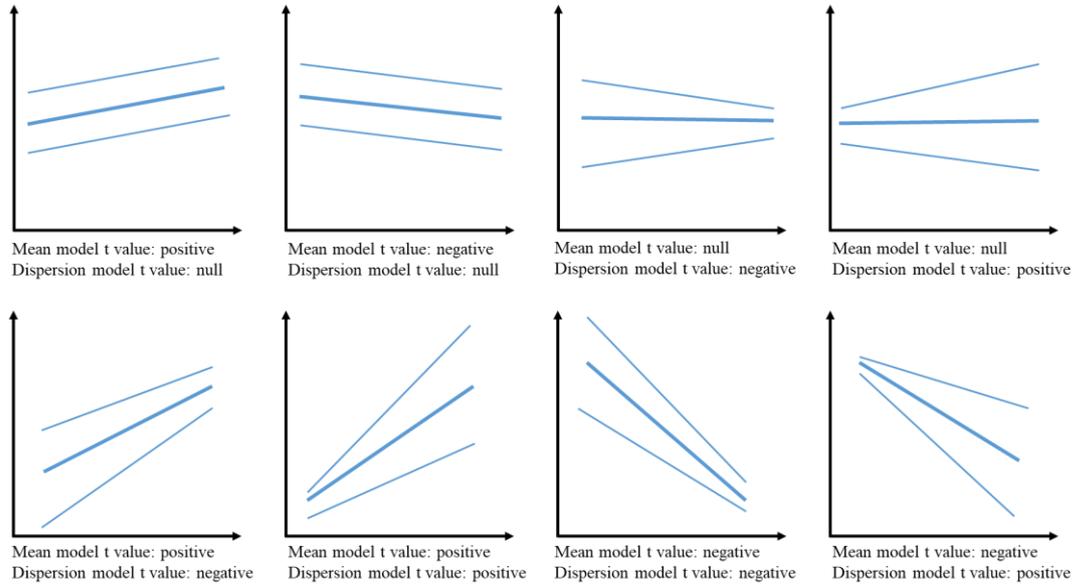

**FigureS2. A general view of relationship between different kinds of data distribution and DGLM.** The thick blue line represents the data mean, and the thin blue line represents the data variance.

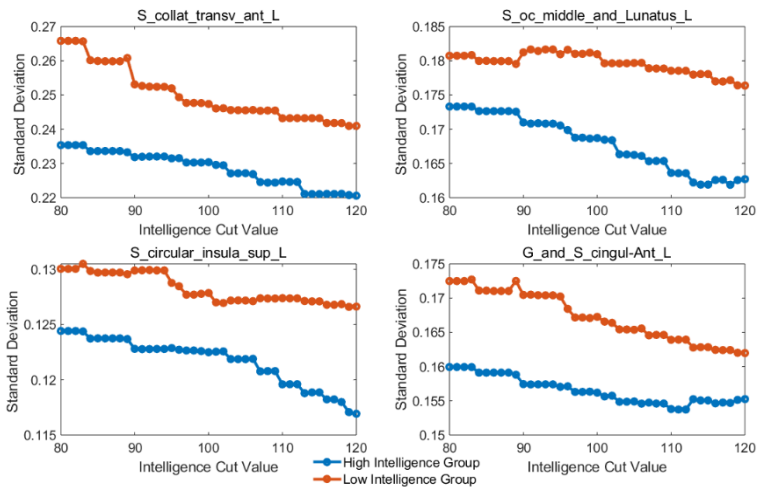

**Figure S3. The standard deviation of four significant regions in variability model of two groups under different intelligence cut value.**

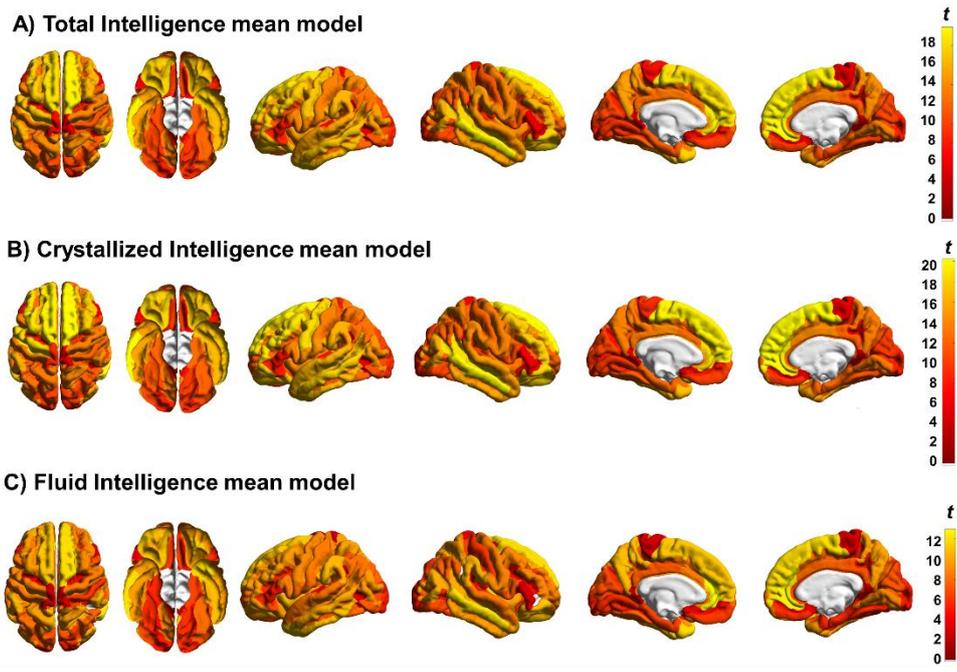

**Fig S4. The mean model results of regions of interest, which the cortical surface area significantly correlation with total, crystallized and fluid intelligence.**

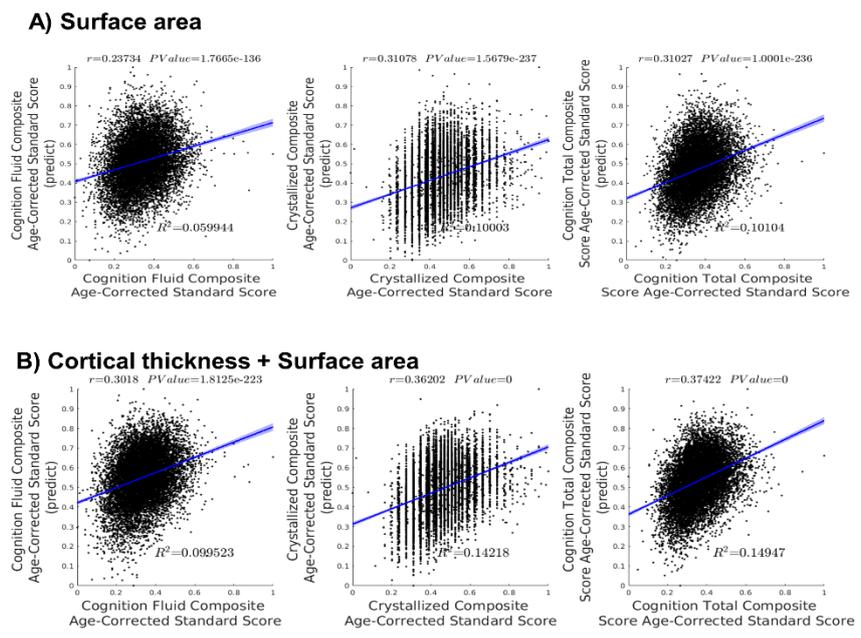

**Fig S5. Scatterplots between elastic net predicted and normalized intelligence scores (y axes) and original normalized values (x axes). A) Surface area were used to predict the g, gC and gF. B) Cortical thickness and surface area were used to predict the g, gC and gF.**

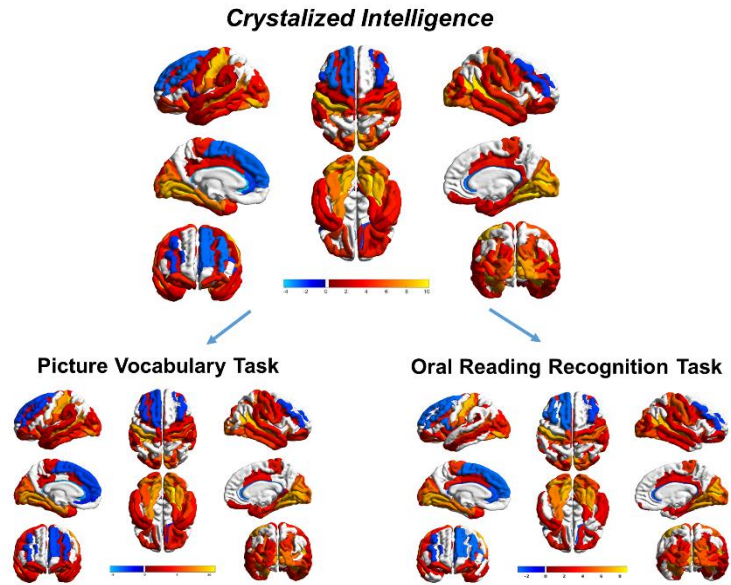

**Fig S6.** The mean model results of regions of interest, which the cortical thickness significantly correlation with gC and its subitems.

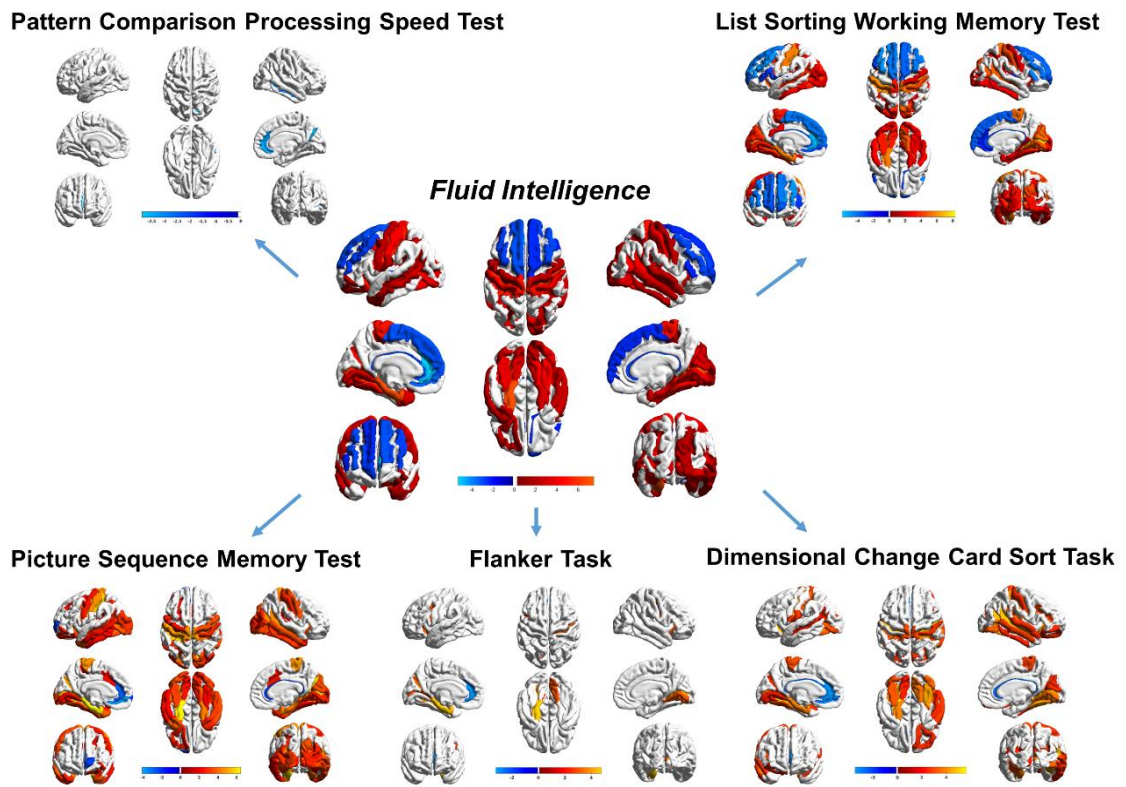

**Fig S7.** The mean model results of regions of interest, which the cortical thickness significantly correlation with gF and its subitems.